# Graphene induced tunability of the surface plasmon resonance

Jing Niu,[1] Young Jun Shin,[1] Youngbin Lee,[2] Jong-Hyun Ahn,[2] and Hyunsoo Yang[1,a)]

[1]*Department of Electrical and Computer Engineering, National University of Singapore, 4 Engineering Drive 3, Singapore 117576, Singapore*

[2]*School of Advanced Materials Science & Engineering, SKKU Advanced Institute of Nanotechnology, Sungkyunkwan University, 300 Cheoncheon, Jangan, Suwon, Gyeonggi, 440-746, Republic of Korea*

Tunability of the surface plasmon resonance wavelength is demonstrated by varying the thickness of $Al_2O_3$ spacer layer inserted between the graphene and nanoparticles. By varying the spacer layer thickness from 0.3 to 1.8 nm, the resonance wavelength is shifted from 583 to 566 nm. The shift is due to a change in the electromagnetic field coupling strength between the localized surface plasmons excited in the gold nanoparticles and a single layer graphene film. In contrast, when the graphene film is absent from the system, no noticeable shift in the resonance wavelength is observed upon varying the spacer thickness.

[a)] e-mail address: eleyang@nus.edu.sg

Graphene consists of a layer of carbon atoms arranged in a honeycomb lattice structure. Since it was first made from graphite by exfoliation in 2004, it has become one of the most popular materials under intense research in condensed matter physics and electronic engineering due to its exceptional electromechanical properties.[1, 2] The high mobility and the low momentum scattering rate of the carriers in graphene make it a technologically important material for high frequency devices. A top gated graphene transistor by chemical vapor deposition (CVD) with a cut off frequency of 155 GHz has been demonstrated recently.[3] In addition, graphene is considered to be a promising candidate to replace indium tin oxide (ITO) as transparent conductive electrodes in optoelectronics applications. Advances in large scale graphene synthesis methods enable the development of a graphene based touch-screen panel.[4, 5] Graphene is expected to have a significant impact in various fields in the foreseeable future.

Localized surface plasmon resonance (LSPR) in conventional systems is a result of collective electron oscillations excited by light in metal nanoparticles. It induces a strong confinement and enhancement of electric fields near the vicinity of metal nanoparticles.[6] The generation of LSPR stimulates various effects such as surface enhanced Raman spectroscopy, plasmon enhanced fluorescence, and surface plasmon enhanced absorption in solar cells.[7-10] The plasmon resonance wavelength is crucial in all of these applications, therefore its tunability to a desired wavelength is greatly beneficial. This can be achieved by controlling the size, the shape, the material of the nanoparticles, and the dielectric constant of the surrounding media.[11] Another effective way of realizing the tunability of LSPR is by the introduction of a metal film in the vicinity of metal nanoparticles. In addition, some other features exhibited by the interaction between a conducting film and metal nanoparticles make it an attractive candidate for nanophotonic applications. For example, the particle-film system functions as an effective optical antenna, capable of localizing the visible radiation to subwavelength dimensions in order to control the interaction between light and matter at the nanoscale.[12] Previously both theoretical and experimental studies proved the tunability of the LSPR wavelength by changing the distance between metal nanoparticles and a metal film due to the change in the coupling strength of the electromagnetic field surrounding the particles and the metal film.[12, 13]

In this letter, we report the tunability of the wavelength of LSPR by varying the distance between graphene and Au nanoparticles. It is estimated that every nanometer of change in the distance between graphene and the nanoparticles corresponds to a resonance wavelength shift of ~12 nm. The nanoparticle-graphene separation changes the coupling strength of the electromagnetic field of the excited plasmons in the nanoparticles and the antiparallel image dipoles in graphene.

A 1 cm × 1 cm single layer graphene thin film, grown by CVD on a copper film is transferred to a transparent borosilicate glass substrate. A Raman spectrum of graphene on a bare borosilicate glass substrate is shown in Fig. 1(a). The G and 2D peaks confirm the presence of graphene with no detectable D peak suggesting the absence of microscopic disorder in graphene. The transmission spectra of a borosilicate glass substrate without and with graphene are shown in the inset of Fig. 1(a). With a graphene layer on top of the glass substrate, the transmittance reduces by ~ 2%, which is comparable with the previously reported opacity of single layer graphene.[14] A layer of Al with a thickness of less than 2 nm is deposited on top of the sample by an electron beam evaporator, followed by natural oxidation under ambient conditions. $Al_2O_3$ films of four different thicknesses are prepared in different areas of the same graphene sample. The thickness of $Al_2O_3$ is estimated to be 0.3, 0.7, 0.9, and 1.8 nm, respectively, by an ellipsometry technique. On top of the $Al_2O_3$ layer, 1.5 nm of Au film is deposited and the scanning electron microscope (SEM) image shows the formation of Au nanoparticles as shown in the upper inset of Fig. 1(b). The Au metal nanoparticles are spheres with a diameter of ~

10 nm. Figure 1(b) illustrates the structure of the sample. The variation of the $Al_2O_3$ thickness shows no noticeable difference in the average size of Au nanoparticles. LSPR is characterized by a UV-visible spectrophotometer in the transmission mode. An unpolarized light beam incident perpendicular to the sample surface is used as the LSPR excitation source.

The transmittance of the bare glass substrate with various thicknesses of $Al_2O_3$ remains at a constant value in the range of the measurement (400-1800 nm) as shown in Fig. 2(a). A similar result is seen for graphene samples with various thicknesses of $Al_2O_3$ shown in Fig. 2(b). After the deposition of Au nanoparticles on top of the $Al_2O_3$ layer, the transmission spectrum is taken for both bare glass samples and graphene samples. An obvious difference in transmittance is observed between samples without and with a graphene film as shown in Fig. 2(c) and (d), respectively. In Fig. 2(c) the LSPR resonance dip of the bare glass substrate shows no noticeable difference despite the variation of the $Al_2O_3$ spacer layer thickness. The resonance wavelength is 569 nm, and the full width at half maximum (FWHM) of the dip is 128 nm. In contrast, for graphene samples in Fig. 2(d), a blue shift of the resonance wavelength is observed with increasing the thickness of the $Al_2O_3$ spacer layer. By varying the spacer layer thickness from 0.3 to 1.8 nm, the resonance wavelength is shifted from 583 to 566 nm. This corresponds to a resonance wavelength shift of ~ 12 nm for every nanometer change in the distance between graphene and nanoparticles, which is a two fold increase in the tunability in comparison to that introduced by the silver nanoparticles and gold thin film system.[12] In addition, the FWHM of the dips decreases as the thickness of $Al_2O_3$ layer increases. Figure 3(a) illustrates the tunability of LSPR wavelength achieved by varying the thickness of the spacer layer between graphene and Au particles. Since the thickness of Al is very small and the deposition is performed at room temperature, $Al_2O_3$ may not be a continuous film. However, it is clear that the shift in the resonance wavelength and our conclusions are not affected because of the flatness of the system, supported by the results in Fig 2.

In order to evaluate the quality of graphene film after the deposition of metals on top of graphene, Raman measurements with a 488 nm excitation laser for different thicknesses of the spacer layer are carried out as shown in Fig 3(b). Although a D peak is present in the spectra, there is no noticeable difference for different thicknesses of the spacer layer. Since the level of defect for different thickness of $Al_2O_3$ is similar, the defects are not the origin of the shifting of the LSPR wavelength. In addition, the graphene film is in the early phase of amorphization according to the model by Ferrari and Robertson.[15] The in-plane correlation length ($L_a$) is estimated to be ~8.7 nm which is well above the limit of conductivity lost of graphene sheet. [16, 17] Therefore, the presence of defects does not affect the functionality of graphene as a conductive film below Au nanoparticles.

The wavelength shift in LSPR is caused by the coupling between the localized electromagnetic field surrounding the metal nanoparticles and the conducting film.[13] Since the incident unpolarized light beam is perpendicular to the sample surface, the electric field has no vertical component with respect to the sample surface, which means only lateral electron oscillations in the metal nanoparticles can be induced. When the distance between metal nanoparticles and the conducting film is small, an antiparallel image dipole will be formed in the conducting film. The presence of an antiparallel image dipole will reduce the internal field in the nanoparticles, which results in a red shift of the resonance wavelength and the interaction between the dipoles will decrease as the spacer layer thickness increases.[12] As a result, the resonance wavelength shows a blue shift with increasing the spacer layer. Therefore, our result infers the formation of laterally oscillating image dipoles in graphene. As proposed theoretically, the laterally oscillating image dipoles introduce less resonance shift as compared to the vertically oscillating image dipoles.[13] To induce the vertically oscillating image dipoles, an electrical field with a component perpendicular to the sample surface must be present. For this purpose, the sample is titled for various angles (30°, 45°, and 60°) in our experimental setup and the transmission spectra are measured. But no additional resonance dip is observed and there is no observable change in the resonance wavelength compared to the case of zero tilting, suggesting that no vertical oscillating dipole is present in our sample.

Theoretical calculation based on dipole approximation has been carried out to compare with the experimental results. A schematic illustration of the structure for calculations is shown in the inset of Fig. 3(c). A gold nanosphere is floating in air above a graphene substrate. The thickness of the substrate is assumed to be semi-infinite. The dielectric constant of graphene is based on an assumption that the optical response of every graphene layer is given by optical sheet conductivity.[18] The dielectric constant of gold is taken from literature.[19] A gold nanosphere is represented by a single dipole. Considering the formation of an antiparallel dipole in the substrate, the polarizability of the gold sphere is given by

$$\alpha = 4\pi a^3 \left( \frac{\varepsilon_1 - \varepsilon_2}{\varepsilon_1 + 2\varepsilon_2} \right) \left[ 1 - \beta \left( \frac{a}{2d} \right)^3 \left( \frac{\varepsilon_1 - \varepsilon_2}{\varepsilon_1 + 2\varepsilon_2} \right) \left( \frac{\varepsilon_3 - \varepsilon_2}{\varepsilon_3 + \varepsilon_2} \right) \right]^{-1} \quad (1)$$

in which $\alpha$ is the polarizability of the gold sphere, $a$ is the radius of the sphere (5 nm), $\beta$ is taken as 2 for the vertical electric field or 1 for the lateral electric field, and $d$ is the distance from the edge of the gold sphere to the substrate surface.[13] $\varepsilon_1$, $\varepsilon_2$ and $\varepsilon_3$ are the dielectric constants of gold, air, and the substrate, respectively. The absorption efficiency $Q_{abs}$ and the scattering efficiency $Q_{sca}$ are given by $Q_{abs} = [k/\pi a^2]\,\mathrm{Im}(\alpha)$ and $Q_{sca} = [k^4/6\pi^2 a^2]|\alpha|^2$, respectively.[13] The summation of these two gives the extinction efficiency, $Q_{ext}$. The calculated result of 1- $Q_{ext}$ , which is proportional to the transmission value, versus wavelength for the vertical electric field is shown in Fig. 3(c) and that of the lateral electric field is shown in Fig. 3(d). Despite of only a single sphere in the calculation, the result is in good agreement with the experimental result, showing a blue shift of the resonance dip as the distance between a sphere and the substrate increases. A small difference in the surface plasmon resonance position between calculation and experiment could be due to the assumptions in calculation such as a semi-infinite substrate and the structure difference between calculation and experiments.

In conclusion, the wavelength of localized surface plasmon resonance depends on the thickness of the $Al_2O_3$ spacer layer between graphene and Au nanoparticles. By increasing the spacer layer thickness, a blue shift is observed in the resonance wavelength. The observed phenomenon is attributed to the formation of oscillating image dipoles in graphene and the calculation result based on dipole approximation shows good agreement qualitatively with the experimental data. This is the first experimental demonstration of LSPR resonance tuning in the nanoparticle-graphene structure.

This work is supported by the Singapore National Research Foundation under CRP Award No. NRF-CRP 4-2008-06.

Figure captions

Figure 1. (a) A Raman spectrum of single layer CVD graphene using an excitation wavelength of 532 nm. The inset of (a) shows the transmisstion data of a borosilicate glass substrate without and with graphene. (b) An illustration of the sample structure. The upper inset in (b) is a SEM image of Au nanoparticles formed on top of $Al_2O_3$. The lower inset in (b) is the cross section view of the device structure.

Figure 2. (a) Transmission spectra of a glass substrate capped with different thicknesses of $Al_2O_3$. (b) Transmission spectra from samples of glass/graphene/$Al_2O_3$. (c) Transmission spectra from samples of glass/$Al_2O_3$/particles. (d) Transmission spectra from samples of glass/graphene/$Al_2O_3$/particles with various thicknesses of $Al_2O_3$. The inset of each figure illustrates the cross section view of the device structure. The arrow in (d) shows a shift in the resonance wavelength.

Figure 3. (a) Dependence of the resonance wavelength and FWHM on the thickness of $Al_2O_3$. (b) Raman spectra of graphene samples after the deposition of $Al_2O_3$ and Au nanoparticles. (c) Calculation results of the surface plasmon resonance wavelength excited by vertical electric fields. (d) Calculation results based on lateral electric fields. The arrow shows a shift in the resonance wavelength. The inset in (c) is the schematic configuration of the structure used for calculation.

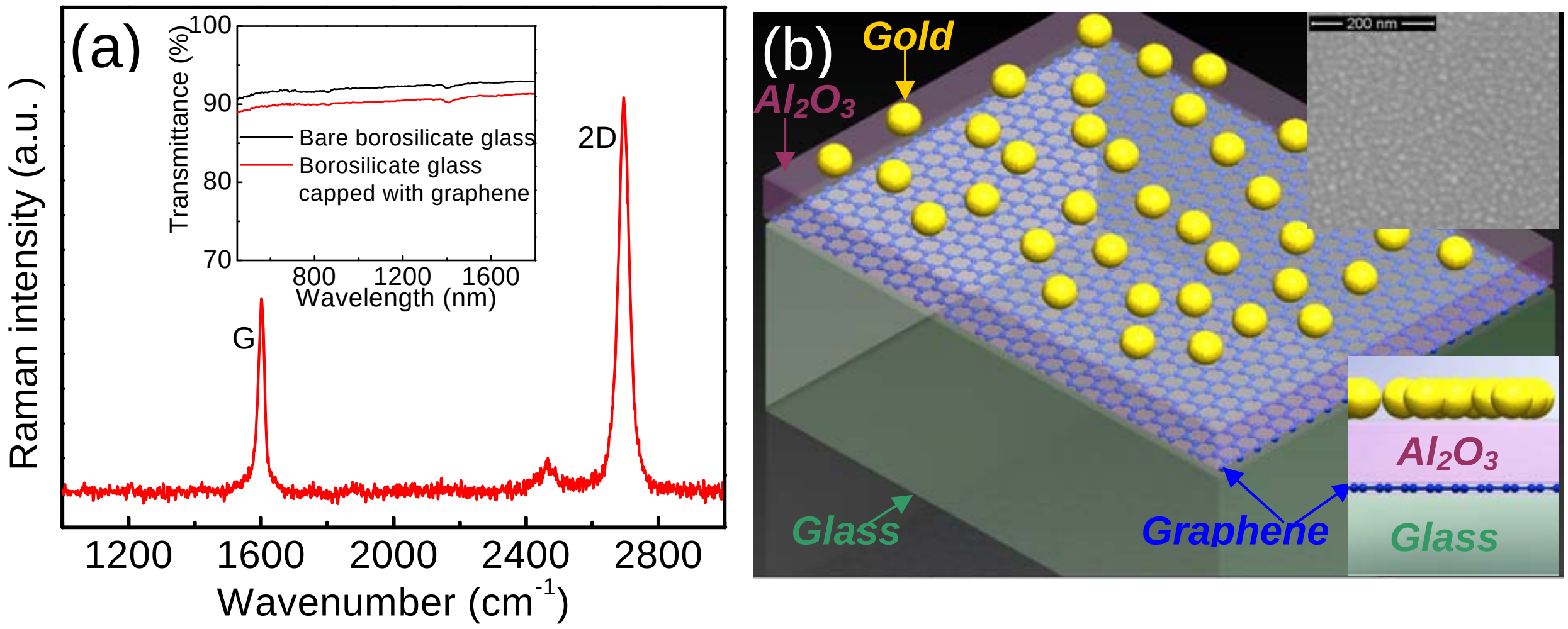

Figure 1

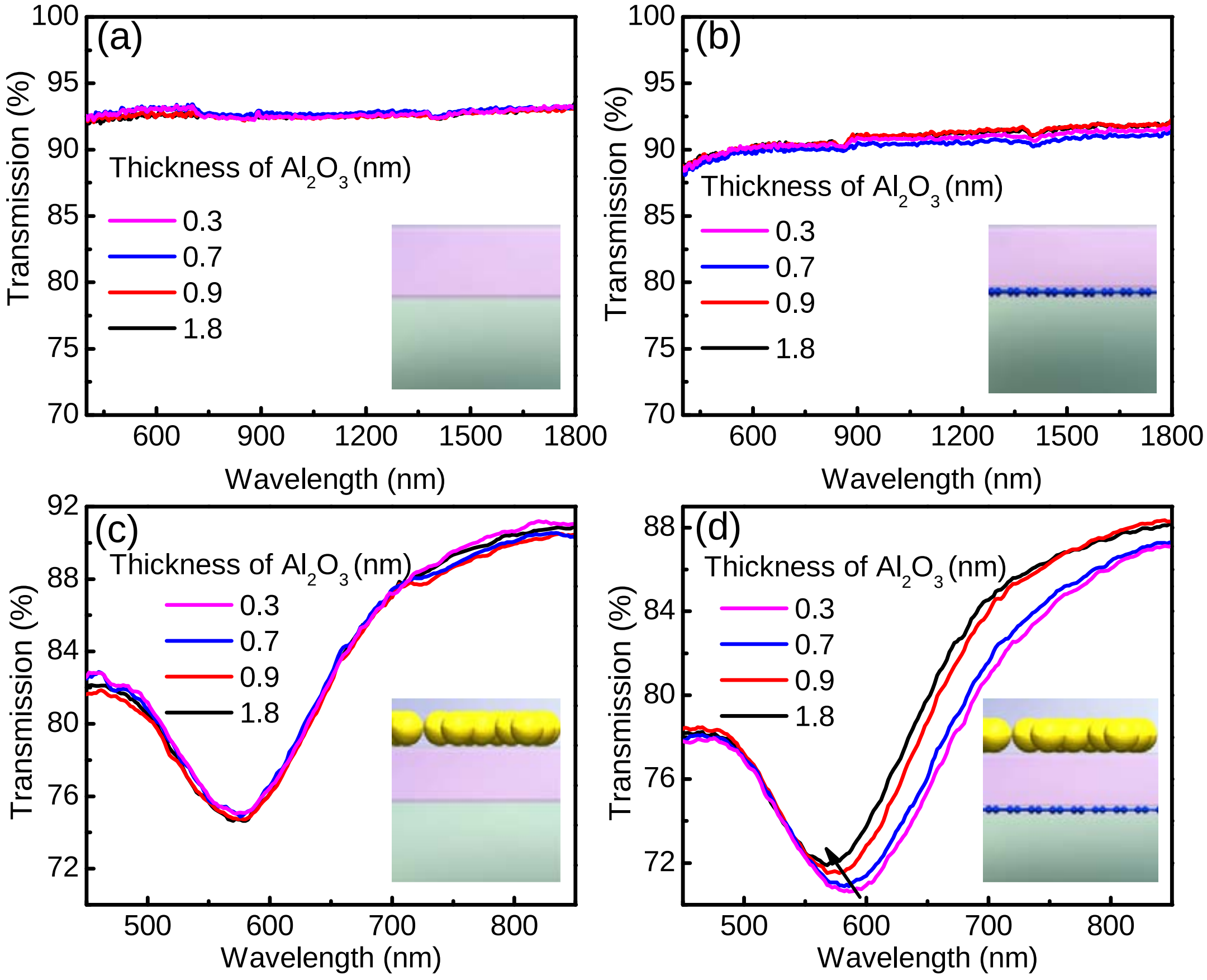

Figure 2.

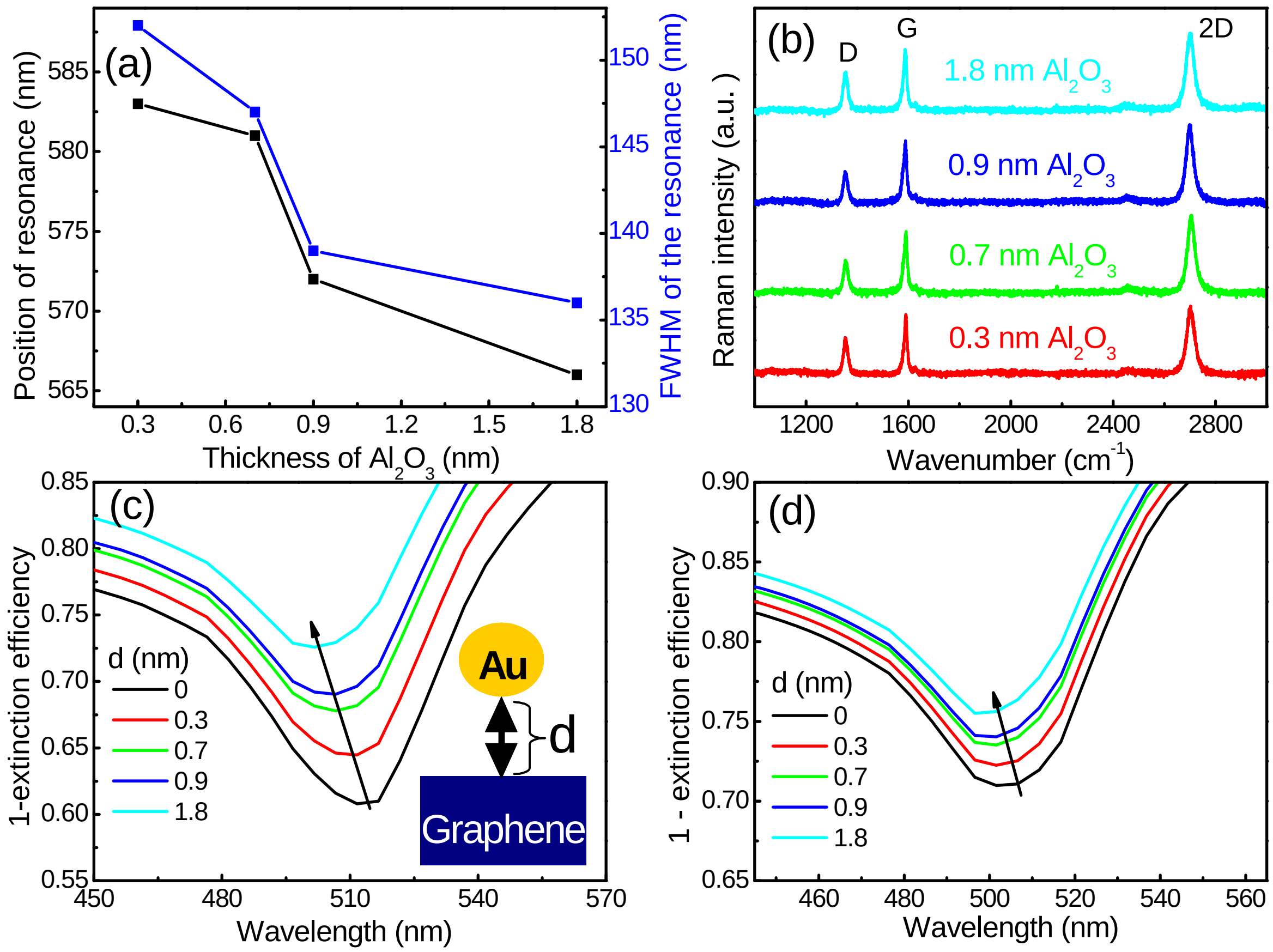

Figure 3.